\documentclass[prc,aps]{revtex4}
\usepackage{epsfig}                                          

\def\bra{\langle} \def\ket{\rangle} 
\def\72Br{$^{72}$Br}
\def\74Br{$^{74}$Br}
\def\78Rb{$^{78}$Rb}

\begin{document}

\title{Projected shell model study of odd-odd $f$-$p$-$g$ shell proton-rich nuclei}
\author{R. Palit$^{1,2}$, J.A. Sheikh$^{3}$, Y. Sun$^{4,5,6,7}$ and H.C. Jain$^{1}$}
\address{$^{(1)}$Tata Institute of Fundamental Research, Mumbai 400 005, India\\
$^{(2)}$Institut f\"ur Kernphysik, Johann Wolfgang Goethe Universit\"at,
D-60486 Frankfurt, Germany\\ 
$^{(3)}$Physik-Department, Technische Universit\"at M\"unchen,
D-85747 Garching, Germany\\
$^{(4)}$Department of Physics and Astronomy, University of Tennessee,
Knoxville, Tennessee 37996\\
$^{(5)}$Department of Physics, University of Notre Dame,
Notre Dame, Indiana 46556\\
$^{(6)}$Department of Physics, Tsinghua University,
Beijing 100084, P.R. China \\
$^{(7)}$Department of Physics, Xuzhou Normal University,
Xuzhou, Jiangsu 221009, P.R. China}

\begin{abstract}
A systematic study of 2-quasiparticle bands of the proton-rich odd-odd nuclei
in the mass $A\sim 70-80$ region is performed using the projected shell model 
approach. 
The study includes Br-, Rb-, and Y-isotopes with N = Z+2, and Z+4. We 
describe the energy spectra and electromagnetic transition strengths in terms 
of the configuration mixing of the angular-momentum projected 
multi-quasiparticle states. Signature splitting and signature inversion in the 
rotational bands are discussed and are shown to be well described. 
A preliminary study of the odd-odd N = Z nucleus, $^{74}$Rb using
the concept of spontaneous symmetry breaking is also presented.
\end{abstract}

\maketitle

\section{Introduction}

Deformed odd-odd nuclei provide valuable information on
the interplay between the collective and the single-neutron and -proton
degrees of freedom in atomic nuclei. 
In general, odd-odd nuclei are difficult to study both experimentally and theoretically
due to the complexity of their low-lying spectra. In comparison to even-even nuclei 
where the pairing correlations favor a particular configuration, in odd-odd nuclei
many configurations are equally probable, and therefore, contribute to the complexity
of the spectra. A detailed experimental analysis for the odd-odd nuclei 
with mass $A\sim 70-80$ has become possible
only in the last few years with the availability of large arrays
of high resolution HPGe detectors. Some extensive measurements of the 
proton-rich odd-odd nuclei 
\cite {70br1,70br2,72br1,74br1,74br2,74rb1,76rb1,78rb1,78rb2,78y1,80y1,82y1} 
have been carried out for the Br-, Rb- and Y-isotopes in laboratories around the world.
The excited states of some of these nuclei have been populated upto high-spins,
e.g. $I \approx 20 \hbar$, by heavy-ion fusion reactions. In each of these 
nuclei several bands have been established. The low-lying levels of 
these bands show very irregular behavior, 
whereas at moderate rotational-frequency, i.e. 
$\hbar\omega \approx 0.5$ MeV, regular rotational bands start developing. 
The stretched $E2$ transition strengths deduced from measured lifetimes of excited 
states indicate higher degree collectivity for the high-spin states. 
Furthermore, a considerable amount of odd-even staggering (signature splitting) in transition 
energies is observed in these bands. In some of them, the staggering 
phase shows an inversion  at spin $I\approx 10$. 

The study of these proton-rich nuclei is not only interesting from the nuclear structure
point of view, but also has important implications in nuclear astrophysics
\cite{rp}. 
Since heavier elements are made in stellar evolution and  explosions, 
nuclear physics, and in particular nuclear structure far from stability, enters 
into the stellar modeling in a crucial way.
It is believed that these proton-rich nuclei near the $N=Z$ line are synthesised
in the rapid-proton capture process (the rp-process) under appropriate 
astrophysical conditions. The X-ray burst is suggested as a possible site.
Recently, the proton-rich odd-odd nuclei
with $A\sim 70-80$ have been studied more rigorously partially due to this astrophysical
interest.
Understanding the structure of the low-lying states in these nuclei is also important 
for the study of Fermi superallowed $\beta^+$ decays \cite{rp3}.  

On the theoretical front, very little effort has been put in the study of
doubly odd nuclei as compared to even-even nuclei in this mass region. 
In an early work, the rotational band structure of doubly odd nuclei in this
mass region has been investigated in $^{76}$Br \cite {kr79} on the basis of 
2-quasiparticle rotor model. More recently, the signature inversion phenomenon 
in these nuclei have been studied in detail via the axially symmetric rotor
plus quasiparticle model \cite {re01}. The 
microscopic calculations based on the Excited-VAMPIR approach \cite{pa1,pa2} 
were performed for some odd-odd nuclei. The large-scale
spherical shell model \cite{cau94} which is quite successful for describing 
the $pf$-shell nuclei, can not be applied for these well-deformed mass-80 
nuclei since it is important to include the $g-$shell. The configuration 
space required for such a study is quite enormous and can not be handled 
by the present state-of-the-art computational facilties.

In recent years, the projected shell model (PSM) \cite{ha95} has become quite 
successful in explaining a broad range of properties of deformed nuclei in 
various regions of the nuclear periodic-table. 
The most striking aspect of this quantum mechanical 
model is its ability to describe the finer details of the high-spin spectroscopy 
data with simple physical interpretations. 
The studies of  odd-odd rare-earth nuclei \cite{ha91} have shown this capability.
Very recently, the PSM has been systematically applied to the even-even 
Kr-, Sr-, and Zr-isotopes of the proton-rich region\cite {pa01}, 
where it has been found that the PSM
results are able to explain most of the experimental observations.
The purpose of the present work is to carry out a similar study 
for the proton-rich, odd-odd 
Br-, Rb- and Y-isotopes with N = Z+2 and Z+4, and with an example of N = Z nucleus.
The physical quantities to be described are energy spectrum and electromagnetic transition
probability. A quantitative comparison of calculated and measured transition 
strengths can provide a stringent test of the model. We would like to mention 
here that the
present study is the first major application of the projected shell model approach
to the medium mass odd-odd nuclei. It is also for the first time that calculations for
the electromagnetic transition strengths in odd-odd nuclei are performed 
within the PSM framework.  

The manuscript is organized as follows:
In the next section, we shall give a brief outline of the model for completeness. This
section also gives expressions for various physical quantities to be discussed 
in this paper.
The results of calculations and comparisons with the experimental
data are presented in Section 3. Finally, a summary is given in 
Section 4. 

\section{The projected shell model}

In the projected shell model approach the basis in which the shell model
Hamiltonian is diagonalised, is chosen in the quasi-particle (qp) space. For
odd-odd nuclei, the ground-band is a two quasi-particle configuration and in
order to describe the structure of a well-deformed odd-odd nuclei,
it is necessary to consider atleast two qp configurations in the basis. 
As a matter of fact, due to the Pauli blocking of levels, quite upto 
high-spins these two qp
configurations are sufficient to describe the physics. Thus,
for low-lying bands of odd-odd nuclei, the quasiparticle configuration space  
consists of a set of 2-qp states  
\begin{equation}
\{ | \phi_\kappa \ket = {a_\nu}^\dagger {a_\pi}^\dagger | 0 \ket \} . 
\label{intrinsic}
\end{equation}
The qp-vacuum $ | 0 \ket $ is determined by diagonalization of a deformed
Nilsson Hamiltonian and a subsequent BCS calculations. This defines the 
Nilsson + BCS qp-basis. 
Here, the quantum numbers for neutron ($\nu$) and proton ($\pi$) run over the
Nilsson single-particle states near the respective Fermi levels. The configuration space is
obviously large in this case compared to the nearby odd-mass nuclei, and usually several 
configurations contribute to the shell model wave function of a state with 
nearly equal weightage. This makes the numerical results very sensitive to the 
shell filling and the theoretical predictions for doubly-odd nuclei become
far more challenging. 

The states $ | \phi_\kappa \ket $ obtained from the deformed Nilsson calculations  
do not conserve rotational symmetry. To  
restore this symmetry, angular-momentum projection technique is applied. The effect of rotation
is totally described by the angular-momentum projection operator and the whole 
dependence of wave functions on spin is contained in the eigenvectors,
since the Nilsson quasi-particle basis is not spin-dependent.  
From each intrinsic state in (\ref{intrinsic}) a band can be generated by 
projection. The interaction 
between different bands with a given spin is taken into account by diagonalising
the shell model Hamiltonian in the projected basis. 

The Hamiltonian 
used in the present work is
\begin{equation}
\hat{H} = \hat{H_0} - \frac{1}{2} \chi \sum_\mu  \hat{Q}_\mu^\dagger 
\hat{Q}_\mu - G_M \hat{P}^\dagger \hat{P} 
- G_Q \sum_\mu \hat{P}_\mu^\dagger \hat{P}_\mu,
\end{equation}
where $\hat{H_0}$ is the spherical single-particle shell model Hamiltonian.
The second, third 
and fourth terms represent quadrupole-quadrupole, monopole-pairing, and 
quadrupole-pairing interactions,
respectively. The strength of the quadrupole-quadrupole force $ \chi $ is 
determined in such a way that the employed quadrupole deformation $ \epsilon_2 $
(see Table I) is same as obtained by the HFB procedure. 
The monopole-pairing force constants $G_M$ used in the calculations are
\begin{equation}
G_M ^\nu = \lbrack 20.25 - 16.20 \frac{N-Z}{A} \rbrack A^{-1}, ~~~~~ 
G_M ^\pi = 20.25 A^{-1} . 
\end{equation}
Finally, the quadrupole pairing strength $G_Q$ is assumed to be proportional to
the monopole strength,
$G_Q = 0.16 G_M$. All these interaction strengths are the same as those in our previous
calculations for the even-even nuclei of the same mass region \cite{pa01,SS01}. 
Thus, we have a consistent description for doubly-even and doubly-odd
nuclear systems. 

Once the projected basis is prepared, we diagonalize the Hamiltonian
in the shell model space spanned by $\hat {P}^{I}_{MK} | \phi_\kappa \ket $.
So, the ansatz for the wave function is given by
\begin{equation}
|\sigma, I M \ket = \sum_{K,\kappa} f^\sigma_\kappa \hat{P}^I_{MK} | \phi_\kappa \ket . 
\end{equation}
Here, the index $\sigma$ labels the states with same angular momentum and $\kappa$
the basis states. $\hat{P}^I_{MK}$ is angular momentum projection operator
and $f^\sigma_\kappa$ are the weights of the basis state $\kappa$. 
This leads to the eigenvalue equation
\begin{equation}
\sum_{\kappa '} (H_{\kappa \kappa'} - E_\sigma N_{\kappa \kappa'} ) f^{\sigma}_{
\kappa'} = 0 ,
\end{equation}
and the normalization is chosen such that
\begin{equation}
\sum_{\kappa \kappa'}f^\sigma_\kappa N_{\kappa \kappa'} f^{\sigma'}_{\kappa'}
= \delta_{\sigma \sigma'}.
\end{equation}                                                      

The angular-momentum-projected wave functions 
are laboratory wave functions and can thus be directly used to compute 
the observables. 
The reduced electric quadrupole transition probability $B(E2)$ from an initial state 
$( \sigma_i , I_i) $ to a final state $(\sigma_f, I_f)$ is given by \cite {su94}
\begin{equation}
B(E2,I_i \rightarrow I_f) = {\frac {e^2} {2 I_i + 1}} 
| \bra \sigma_f , I_f || \hat Q_2 || \sigma_i , I_i\ket |^2 .
\end{equation}
In the calculations, we have used the effective charges of 1.6e for protons 
and 0.6e for neutrons. 
The reduced magnetic dipole transition probability
$B(M1)$ is computed by 
\begin{equation}
B(M1,I_i \rightarrow I_f) = {\frac {\mu_N^2} {2I_i + 1}} | \bra \sigma_f , I_f || \hat{\mathcal M}_1 ||
\sigma_i , I_i \ket | ^2 , 
\end{equation}
where the magnetic dipole operator is defined as  
\begin{equation}
\hat {\mathcal {M}}_{1}^\tau = g_l^\tau \hat j^\tau + (g_s^\tau - g_l^\tau) \hat s^\tau . 
\end{equation}
Here, $\tau$ is either $\nu$ or $\pi$, and $g_l$ and $g_s$ are the orbital and the spin gyromagnetic factors, 
respectively. 
In the calculations
we use for the $g_l$ the free values and for $g_s$ the free values damped by a 0.85 factor
\begin{equation}
g_l^\pi = 1, ~~~ 
g_l^\nu = 0, ~~~   
g_s^\pi =  5.586 \times 0.85, ~~~ 
g_s^\nu = -3.826 \times 0.85.
\end{equation}
Since the configuration space is large enough we do not use any core 
contribution.
More concretely, the reduced matrix element $\hat {\mathcal {O}}$ ($\hat {\mathcal {O}}$ is either
$\hat {Q}$ or $\hat {\mathcal {M}}$) is expressed by
\begin{eqnarray}
\bra \sigma_f , I_f || \hat {\mathcal {O}}_L || \sigma_i , I_i\ket   &  = &
\sum_{\kappa_i , \kappa_f} {f_{I_i \kappa_i}^{\sigma_i}} {f_{I_f \kappa_f}^{\sigma_f}}
\sum_{M_i , M_f , M} (-)^{I_f - M_f}  
\left( \begin{array}{ccc}
 I_f & L & I_i \\
-M_f & M & M_i 
\end{array} \right) \nonumber \\
 & & \bra \phi_{\kappa_f} | {\hat{P}^{I_f}}_{K_{\kappa_f} M_f} \hat {\mathcal {O}}_{LM}
\hat{P}^{I_i}_{K_{\kappa_i} M_i} | \phi_{\kappa_i} \ket  
\end{eqnarray}


\section{Results and discussions}

For N = Z+2, Z+4 isotopes of odd-odd Br, Rb, and Y, 
the $\pi g_{9/2} \otimes \nu g_{9/2}$, 2-qp configuration
has been firmly established for the yrast region. 
This configuration has been observed up to or above
10 MeV excitation energy in these nuclei. Therefore, 
we first discuss the calculations for these nuclei 
in the following three subsections. 
The discussion includes 
the band diagrams, transition energies and electromagnetic transition strengths
(the B(M1) and B(E2) values) of each of the above mentioned isotopes. 

The odd-odd N = Z nuclei of this mass region appear to have a different
structure as compared to the neighbouring nuclei. 
Experimentally, the odd-odd N = Z nuclei are far more difficult to 
study since they lie further
away from the stability valley, approaching the proton drip-line.
For the few measured cases \cite{70br1,70br2,74rb1}, 
both the low-spin and high-spin states exhibit distinct properties.  
The structure with the presence of the isospin, T = 1 and T = 0 bands 
can be further complicated by the prolate-oblate shape co-existence at low spin
for many nuclei.
Therefore, we preferred to choose the experimentally known N = Z nucleus
$^{74}$Rb \cite{74rb1} which has large prolate deformation even in ground 
state. Separate description of T = 1 and T = 0 bands was given 
to show the applicability of the simple model based on schematic Hamiltonian
to this extreme case. 
However, the present version of PSM does not treat explicitly
the isospin degree of freedom to calculate the interaction between T= 1 and 
T = 0 bands, giving a constraint for the actual descriptin of the structure.


The PSM calculations proceed in two steps. First, an optimum set of deformed
basis is constructed from the standard Nilsson model. The Nilsson 
parameters are taken from Ref. \cite {su00} and the calculations are performed
by considering three major shells (N = 2, 3 and 4) for both neutrons and protons.
This basis is large enough so that all nucleons in the f-p-g shells are active. 
The basis deformation $\epsilon_2$ used for each nucleus is given in Table I.
These values are taken either according to experimental information, if available, or from
theoretical calculations. We emphasize that 
unlike the cranking mean field approaches, the deformation 
parameters used as an input to the PSM calculations need not correspond exactly 
to the true nuclear deformation. This is because of the shell-model nature of the PSM:
The deformed single-particle
states serve solely as a way to truncate shell-model basis. All observable 
properties in the PSM calculations are determined by the many-body wave-functions
obtained by diagonalising the shell model Hamiltonian. In principle, the larger the deviation
of the initial deformation from the true one, the bigger will be the 
configuration space for an actual description of the nuclei. In the present calculations, 
the space is truncated by the inclusion of the states within an energy window of 3.5 MeV 
around the Fermi surface. This determines the size of the basis space, 
$ | \phi_\kappa \ket $ in Eq. (\ref{intrinsic}), of the order of 30. In the second
step, these basis states are projected to good angular momentum states, and
the projected basis is then used to diagonalize the shell model Hamiltonian.
The diagonalization gives rise to the energy spectra, and the transition strengths
are subsequently calculated using the resulting wave functions. 


\subsection{Band diagram}

For the deformation values given in Table I, 
the $[440]\frac
{1}{2},~[431]\frac{3}{2},~[422]\frac{5}{2}$ orbitals for protons and 
neutrons lie near the Fermi 
surfaces in the nuclei under consideration. 
To describe the low-lying structure of these nuclei, the 2-qp 
states based on these orbitals (with $K = K_\nu \pm K_\pi$) should be taken into account.
A representative band diagram for $^{74}$Br,
containing some unperturbed rotational bands,
is shown in Fig. 1 to see the underlying structure. Note that only the important
configurations are displayed in the band diagram,
although many more others are included in the calculations.

Out of the all unperturbed bands taken in the diagonalization, only few of 
them show energy staggering between odd- and even-spin states (the so-called signature splitting). 
As can be seen in Fig. 1, these bands have the zigzag behavior as a function of spin.  
This occurs for the bands with configurations 
involving low-K orbitals. It was noticed in Ref. \cite {ha91} that 
unlike in the particle rotor model, the configurations based on orbitals with
$K \neq \frac{1}{2}$ also contribute to the energy staggering.
The unperturbed bands which show staggering influence the yrast states through 
the band mixing, and thus cause the signature dependence in the yrast band (shown as filled circles in
Fig. 1).
It is seen from Fig. 1 that the lowest $K=4$ band has only small staggering.
The clear signature splitting at around spin $I=10$ come mainly from the mixing of 
the 2-qp band based on 
$( \nu [431]\frac{3}{2} \otimes \pi  [431]\frac{3}{2} )$. 
Two more pairs of bands 
with 
K = 1, 2 based on $( \nu [431]\frac{3}{2} \otimes \pi [440]\frac{1}{2} )$ 
and 
K = 1, 0 based on $( \nu [440]\frac{1}{2} \otimes \pi [440]\frac{1}{2} )$ 
have much stronger signature splitting. They dive down to the yrast region at higher spins, and therefore
bring strong energy staggering to the yrast band. It is thus the configuration mixing that produces 
nicely the observed signature splitting as one will see in Fig. 2 below.

Interestingly, the current PSM calculations indicate that the phase of the energy staggering may change
at certain spins. In Fig. 1, two of the above mentioned 
bands with K = 0 based on $( \nu [440]\frac{1}{2} \otimes \pi [440]\frac{1}{2} )$
and K = 1 based on $( \nu [431]\frac{3}{2} \otimes \pi [440]\frac{1}{2} )$ show 
an inversion in the staggering phase (the so-called signature inversion). 
The inversion appears at $I=6$ for the former,
and $I=10$ for the latter. Note that this is
a signature inversion occurring within one unperturbed band, without involving any dynamic mixture with
other bands. Hara and Sun called this kind of inversion a  
self-inversion, to distinguish this from an inversion involving two bands
with mutually opposite staggering phases \cite {ha91}. 
The phenomenon of self-inversion predicted in Ref. \cite {ha91} has not been
confirmed experimentally. 
Here, the band with self-inversion (emphasized by a solid bold curve at $I=8-11$ in Fig. 1) 
lies at very low excitations and the
inversion appears at spin $I\approx 10$. 
This is a promising candidate that may explain the observed signature inversion 
in the rotational band in $^{74}$Br \cite{74br1} and other nuclei in this mass region. 

\subsection{Transition energies}

After diagonalization at each spin, the lowest energies 
give the energies of the yrast states. The energy differences between adjacent
spin states are compared with the measured transition energies in Fig. 2
for $^{72,74}$Br, $^{76,78}$Rb and $^{80,82}$Y.
The characteristic feature of the six nuclei 
is the exhibition
of a clear energy staggering between the odd- and even-spin states, 
as seen in the $E(I)-E(I-1)$ plot in Fig. 2. 
It is called signature splitting because the odd- and even-spin states
can be classified as two groups specified by the signature quantum number. 
In addition, an inversion in the staggering phase is observed 
in most of the nuclei. Signature inversion in the mass-80 region was discussed in Ref. \cite{re01}.  

As can be seen from Fig. 2, 
for all of the nuclei the agreement between the calculation and experiment is quite
satisfactory above $I \approx 10 \hbar$. The energy splitting at higher spins
is well reproduced in all the cases, indicating that the important influence on the yrast band
from the low-K 
components of the $g_{9/2}$ valance neutrons and protons are correctly accounted for by the configuration mixing.

However, the calculation does not reproduce the signature inversion observed at low spins in 
$^{74}$Br, $^{76,78}$Rb and $^{80,82}$Y. This is rather unfortunate because the mechanism
that may cause the inversion 
is clearly present in the model. In this medium mass region with $N\sim Z$, rotational bands with
mutually opposite signature dependence can not co-exist near the yrast region, and therefore, the only
possibility  
to explain the observed signature inversion is through the self-inversion. In Ref. \cite{ha91}, the favorite  
condition for this to happen was given to nuclei where bands based on configurations 
with K = 0 $( \nu K=\frac{1}{2} \otimes \pi K=\frac{1}{2} )$
and K = 1  $( \nu K=\frac{3}{2} \otimes \pi K=\frac{1}{2} )$ lie low in energy.
$^{74}$Br fulfils these conditions. The failure that our final results do not reproduce the signature inversion 
may indicate that the band mixing is not strong enough to bring the desired
feature of self-inversion into the yrast band. One may generally argue that there should be
an interaction that particularly pushes the unperturbed band with self-inversion further down to affect the
yrast states more strongly. 
Obviously, this force is absent in the schematic Hamiltonian employed in the current model.

\subsection{Electromagnetic transition strengths}

The experimental data from lifetime measurements performed 
for $^{74}$Br, $^{78}$Rb and $^{82}$Y \cite {74br1,74br2,78rb1,78rb2,82y1} 
have separately provided the B(M1) and B(E2) values. For $^{72}$Br and 
$^{76}$Rb, no lifetime measurements have been performed: 
the ratio B(M1)/B(E2) could be obtained from the measured intensity
of transitions. In the present section, these data are compared with our calculations.  

The experimental B(M1) values for $^{74}$Br, $^{78}$Rb and $^{82}$Y are  
plotted in Fig 3. In all the three nuclei, staggering between the odd-
and even-spin states is clearly present. This variation in the B(M1) pattern is directly 
compared with the calculated transition strengths. Except for a few particular states,
the calculation achieves a quantitatively agreement with data. In contrast to the transition
energies, no inversion in the staggering
phase is obtained in both theory and experiment. The measured B(E2) values
for $^{74}$Br, $^{78}$Rb and $^{82}$Y are plotted in Fig. 4. 
The global trend of the B(E2) curve as a function of spin is qualitatively reproduced by the calculation
as shown in the Fig. 4, although some deviations are seen at high spins. 

In Fig. 5, the calculated B(M1)/B(E2) values are compared with
the experimental data wherever available for $^{72}$Br, $^{76}$Rb and $^{80}$Y.
A staggering pattern is again observed in experimentally deduced B(M1)/B(E2)
ratios for those levels for which cascade M1 transitions have been observed.   
Indeed, in $^{72}$Br B(M1)/B(E2) values for the 11$^{(+)}$ and 13$^{(+)}$ levels 
are $\approx~10~
\mu_n^2/(eb)^2 $, while for 10$^{(+)}$ level its value is only 0.03(1) 
$\mu_n^2/(eb)^2 $ \cite {72br1}. These features are quite nicely reproduced 
by the calculation.
The experimental ratio B(M1)/B(E2) in $^{76}$Rb were
determined from the measured intensity of transitions reported in Ref.
\cite {76rb1}. 
The B(M1)/B(E2) values in $^{76}$Rb are generally  
smaller compared
to the ratio observed in $^{72}$Br. This fact reflects the smaller contribution
from the lower K orbitals for $^{76}$Rb.  
Unfortunately, in 
case of $^{80}$Y no intensities for different transitions in the positive 
parity band is
available to allow a determination of the B(M1)/B(E2) ratios experimentally.
The calculated values of B(M1)/B(E2) ratio are small  
compared to those of $^{72}$Br and $^{76}$Rb. 

\subsection{The N = Z nucleus $^{74}$Rb}

In Ref. \cite{74rb1}, two rotational bands were observed in the N = Z nucleus $^{74}$Rb. 
In contrast to the
above discussed N = Z+2 and Z+4 nuclei, the ground-state band of $^{74}$Rb does not look like
a rotational band having the usual 2-qp
structure. 
Instead, the transition energies of the ground state band in $^{74}$Rb 
show similarities to those in even-even nuclei. In fact,
they have been interpreted \cite{74rb1} as being formed from the T = 1 isobaric analog
states of $^{74}$Kr with pairing correlations based on T = 1 neutron-proton pairs. 
At higher rotational frequency, a T = 0 rotational band becomes energetically favored
over the T = 1 ground state band. 
   
As a full description for these observations is not possible in the present 
version of the PSM since the physics is beyond what the model space 
in (\ref{intrinsic}) can provide, we tried to describe separately T = 0 and 
T = 1 bands.  
Using the concept of spontaneous
breaking of the isospin invariance, in ref. \cite{FS99} it was argued 
that the neutron-proton interaction can be effectively considered in a theory with only
neutron-neutron and proton-proton pairings, since one has a freedom 
to choose a one particular direction in the isospin space. 
One thus ends up with the understanding that the T = 1 ground state band 
in an odd-odd N = Z  nucleus may be approached
by a 0-qp state with nn and pp pairings and the T = 0 rotational band has a structure
similar to the 2-qp states \cite{FS99}.  

The two neighboring even-even N = Z nuclei $^{72}$Kr and $^{76}$Sr have a large
deformation with $\epsilon_2 \approx 0.36$ \cite{pa01,SS01}. 
It is thus reasonable to perform a calculation for $^{74}$Rb with this deformation.
At the deformation $\epsilon_2 = 0.36$, the [422]$\frac{5}{2}$ orbital is the closest  
to the respective neutron and proton Fermi levels in $^{74}$Rb. Thus, the band based on the 2-qp
state of $K=5$ ($\nu$[422]$\frac{5}{2} \otimes \pi$[422]$\frac{5}{2}$) is the lowest in energy
among all the 2-qp states (see Fig. 6). 
At higher spins, bands consist of smaller $K$-states $[440]\frac{1}{2},~ [431]\frac{3}{2}$, 
which show stronger signature splitting, mix with the $K=5$ band through
configuration mixing. Therefore, the mixed band follows the signature phase 
that favors the odd-spin states. This explains the absence of even spin states 
in the experimental spectra at the high spin region \cite{74rb1}.  

In Fig. 7, the two observed rotational bands in $^{74}$Rb are compared with 
the calculations in an $E(I) - E(I-2)$ plot.
It is seen that the calculated results fit the data reasonably well. This indicates that
the physical understanding of ref. \cite{FS99} on the band structure in an
odd-odd N=Z nucleus is correct. However, in order to understand the interplay
between the T=0 and 1 bands, it is necessary to include the neutron-proton
pairing in the shell model Hamiltonian.

\section{Summary}

In the present manuscript, we have 
performed a systematic study for the positive parity yrast bands of 
odd-odd nuclei $^{72,74}$Br, $^{76,78}$Rb and $^{80,82}$Y using 
the projected shell model approach.   
Furthermore, an N = Z nucleus $^{74}$Rb has also been studied. 
We discussed many features observed in these nuclei using 
the angular-momentum projected 2-quasiparticle states 
with the employment of a simple quadrupole-quadrupole + monopole-pairing + 
quadrupole-pairing Hamiltonian. The neutron-proton interaction is present only 
in the particle-hole channel. We have pointed out the successes and inadequacies of
the projected shell model approach in explaining the odd-odd nuclei with mass 
$A\sim 70-80$. The main feature that both transition energies and
electromagnetic transition strengths show a staggering pattern has been 
successfully described by the model in terms of configuration mixing. 
In general, a good agreement between the experimental data and calculated 
values has been found in the high-spin region, but there are discrepancies in 
the low-spin domain.   
This indicates that in the high-spin region, the nuclei studied acquire a stable
deformation, which is the basic idea in the development of the projected 
shell
model approach. The disagreement in the low-spin region can be attributed to 
the shape-coexistence and the vibrational degree of freedom, which are not 
considered in the present approach.

Finally, we  conclude with a note that a qualitative description of the
band structures observed in odd-odd N=Z can be obtained by using the 
existing PSM approach and the concept of spontaneous isospin-symmetry 
breaking. However, for a detailed study of the N = Z systems, the 
current model needs to be extended with angular-momentum projection 
on a quasiparticle basis that allows for a mixing of proton and 
neutron single-particle states. The work along these lines is in 
progress and will be reported in the near future.


\newpage

\begin{table}
\begin{center}
\caption{Parameters used in the PSM calculations}
\begin{tabular}{|c|c|c|} \hline
Nuclei & $\epsilon_2$ & G$_Q$ \\ \hline
$^{72}$Br & 0.295 & 0.16  \\
$^{74}$Br & 0.295 & 0.16  \\
$^{74}$Rb & 0.360 & 0.16  \\
$^{76}$Rb & 0.290 & 0.16  \\
$^{78}$Rb & 0.273 & 0.16  \\
$^{80}$Y  & 0.311 & 0.16  \\
$^{82}$Y  & 0.255 & 0.16  \\ \hline
\end{tabular}
\end{center}
\end{table}

\newpage
\begin{figure}
\begin{center}
\epsfxsize = 11.0 cm \epsfysize = 11.0 cm \epsffile{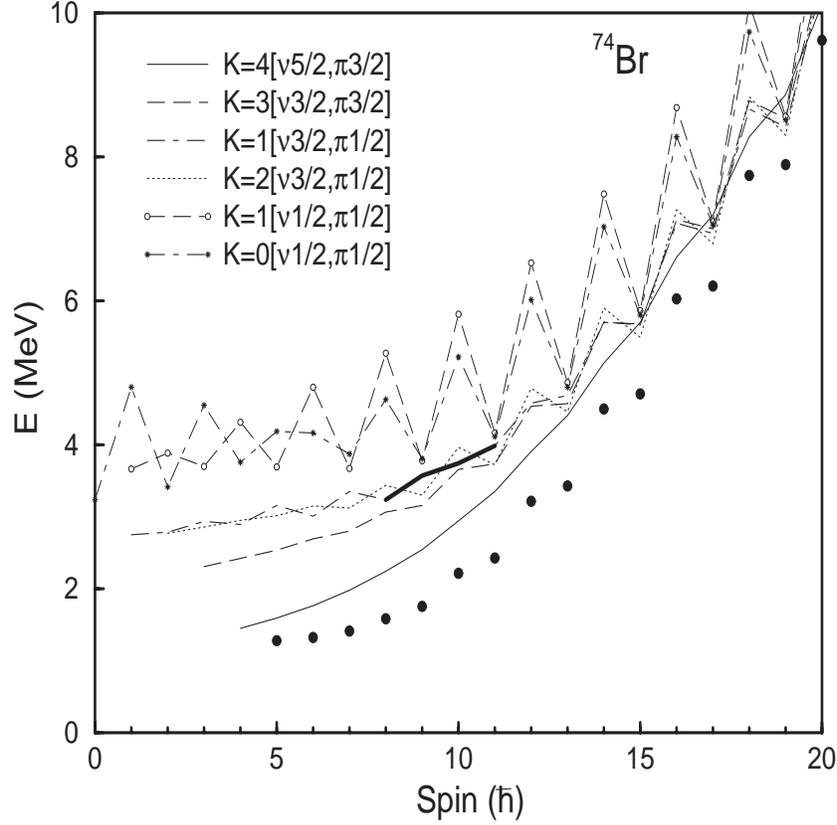}
\caption{A representative band diagram for $^{74}$Br. 
[For other isotopes, relative positions of the bands are different, reflecting
different shell fillings.] Filled circles are the lowest states obtained after
configuration mixing at each spin. Notice in particular the solid, bold curve
at spin $I=8-11$, which shows a self-inversion in that band. }
\end{center}
\end{figure}

~
\begin{figure}
~
\begin{center}
\epsfxsize = 12.0 cm \epsfysize = 12.0 cm \epsffile{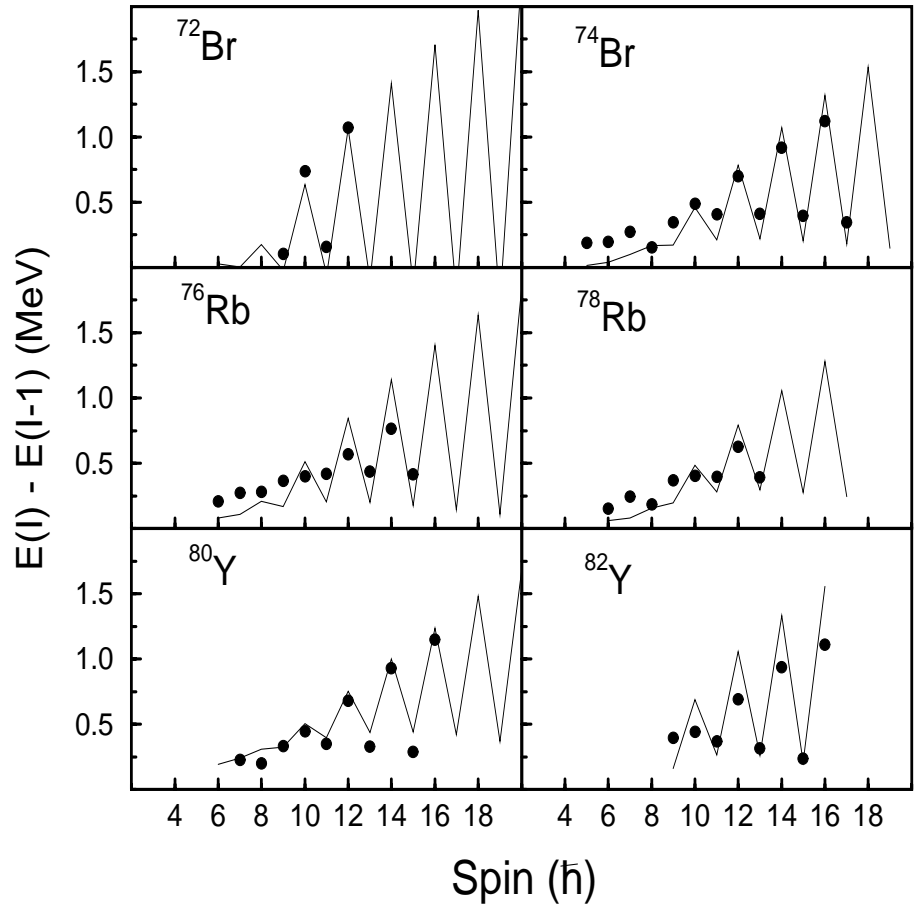}
~
\vspace{1.0cm}
\caption{Comparison of calculated transition energies (solid line) with 
experimental values (filled circle) for the positive parity bands of Br (data from
\protect\cite{72br1,74br1,74br2}), Rb (data from \protect\cite{76rb1,78rb1,78rb2})  
and Y (data from \protect\cite{80y1,82y1}) isotopes.}
\end{center}
\end{figure}

~
\vspace{1.5cm}
\begin{figure}
~
\begin{center}
\epsfxsize = 12.0 cm \epsfysize = 12.0 cm \epsffile{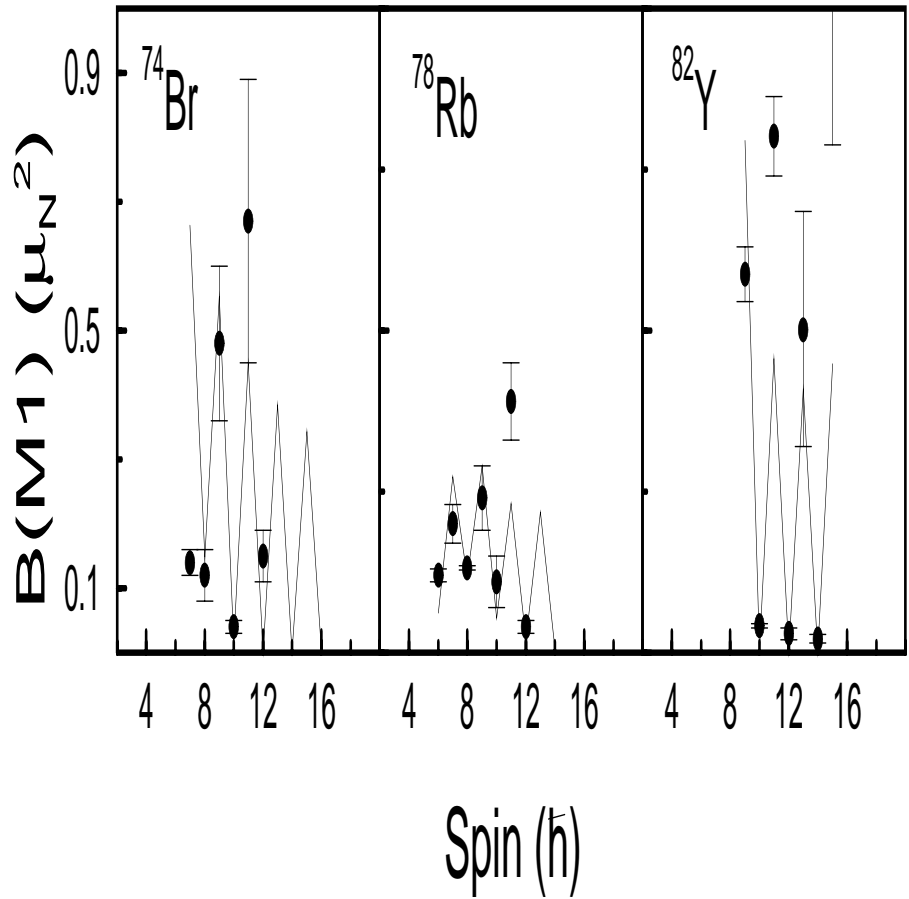}
\caption{ 
Comparison of calculated B(M1) strengths (solid line) 
with the 
measured values (filled circle with error bar) for $^{74}$Br, $^{78}$Rb and $^{82}$Y isotopes.
Data are taken from \protect\cite{74br1,74br2,78rb1,78rb2,82y1} }
\end{center}
\end{figure}

~
\begin{figure}
\begin{center}
\epsfxsize = 12.0 cm \epsfysize = 12.0 cm \epsffile{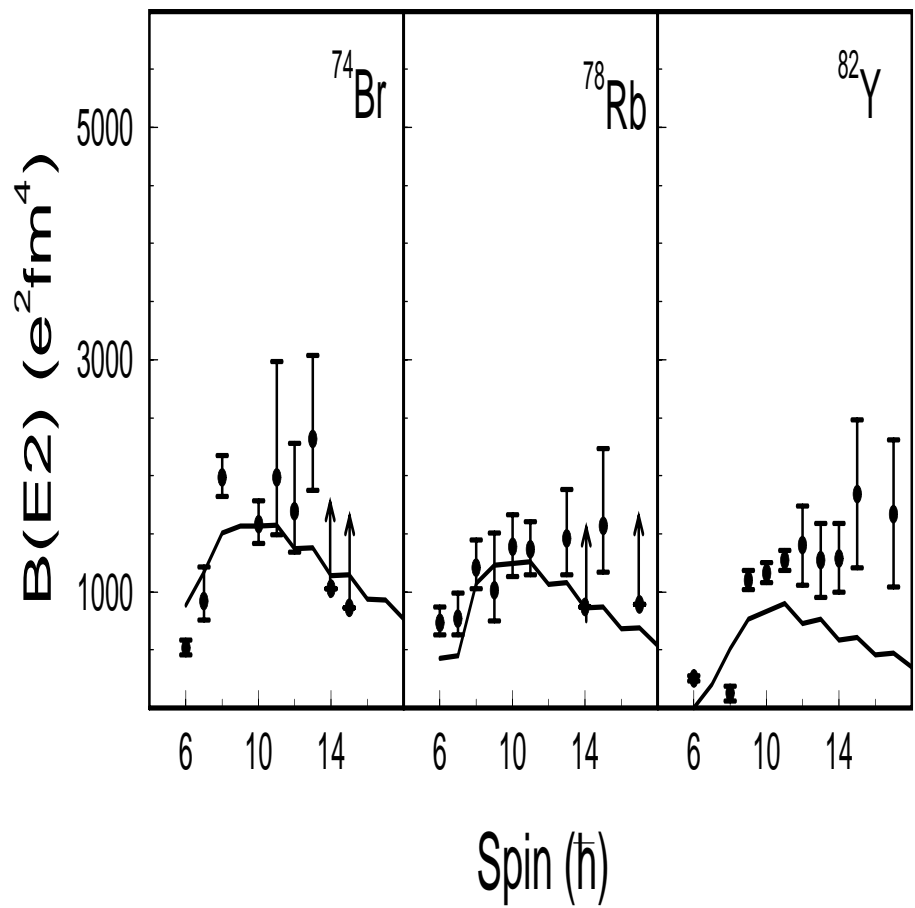}
\caption{Comparison of calculated B(E2) strengths (solid line)
with experimental values (filled circle with error bar) for $^{74}$Br, $^{78}$Rb 
and $^{82}$Y isotopes. Data are taken from \protect\cite{74br1,78rb1,78rb2,82y1}.}
\end{center}
\end{figure}

~
\vspace{1.5cm}
\begin{figure}
~
\begin{center}
\epsfxsize = 12.0 cm \epsfysize = 12.0 cm \epsffile{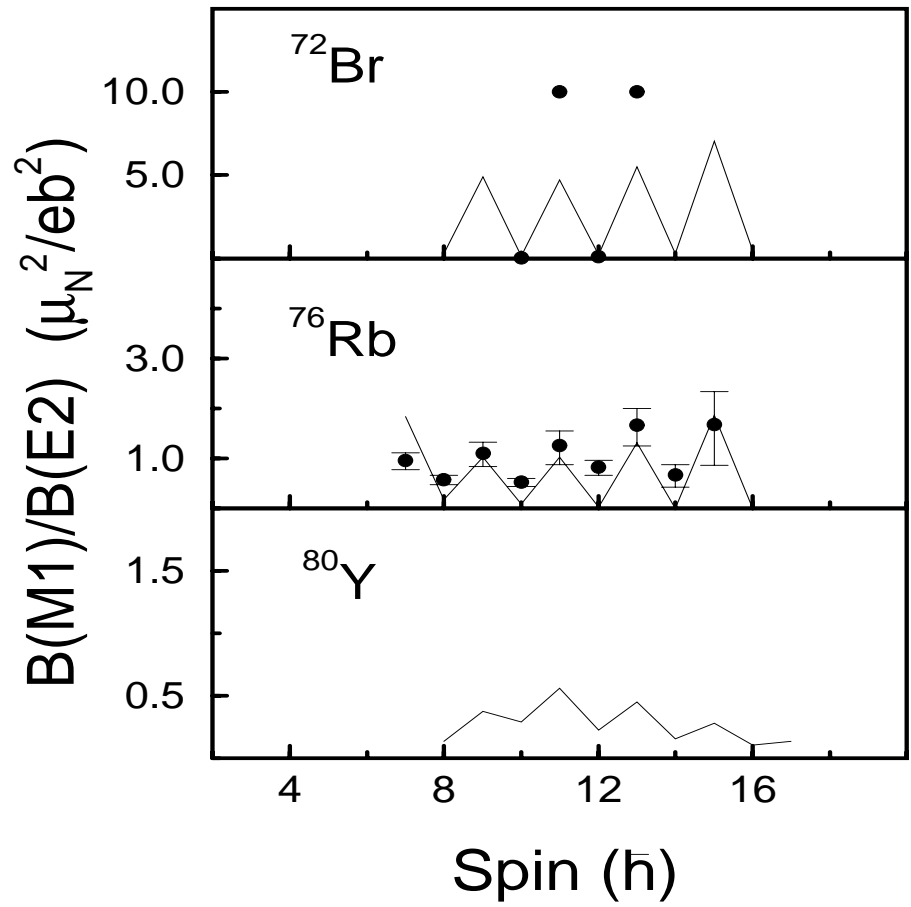}
\caption{Comparison of calculated ratio of B(M1) to B(E2)
with experimental values (filled circle with error bar) for 
$^{72}$Br, $^{76}$Rb and $^{80}$Y isotopes wherever data is available. 
Data are taken from \protect\cite{72br1,76rb1,80y1}.}
\end{center}
\end{figure}

\begin{figure}
\begin{center}
\epsfxsize = 12.0 cm \epsfysize = 12.0 cm \epsffile{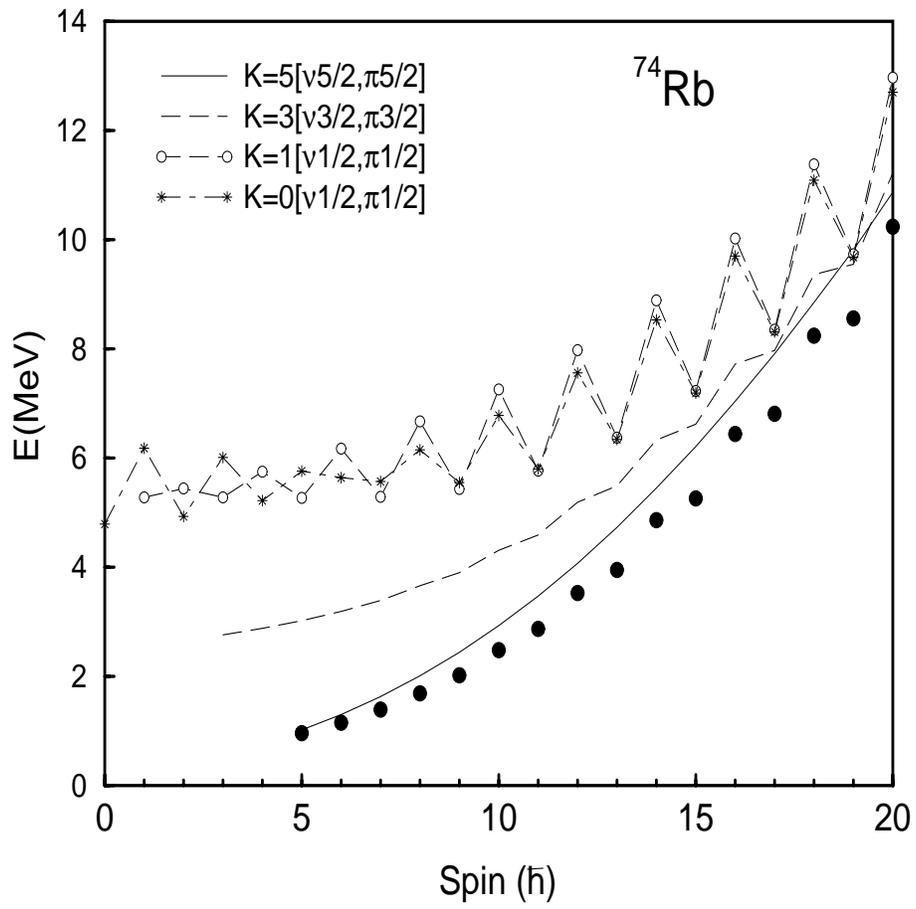}
\caption{Band diagram with 2-qp configurations 
for $^{74}$Rb. Filled circles are the lowest states obtained after
configuration mixing at each spin.}
\end{center}
\end{figure}

\begin{figure}
\begin{center}
\epsfxsize = 12.0 cm \epsfysize = 12.0 cm \epsffile{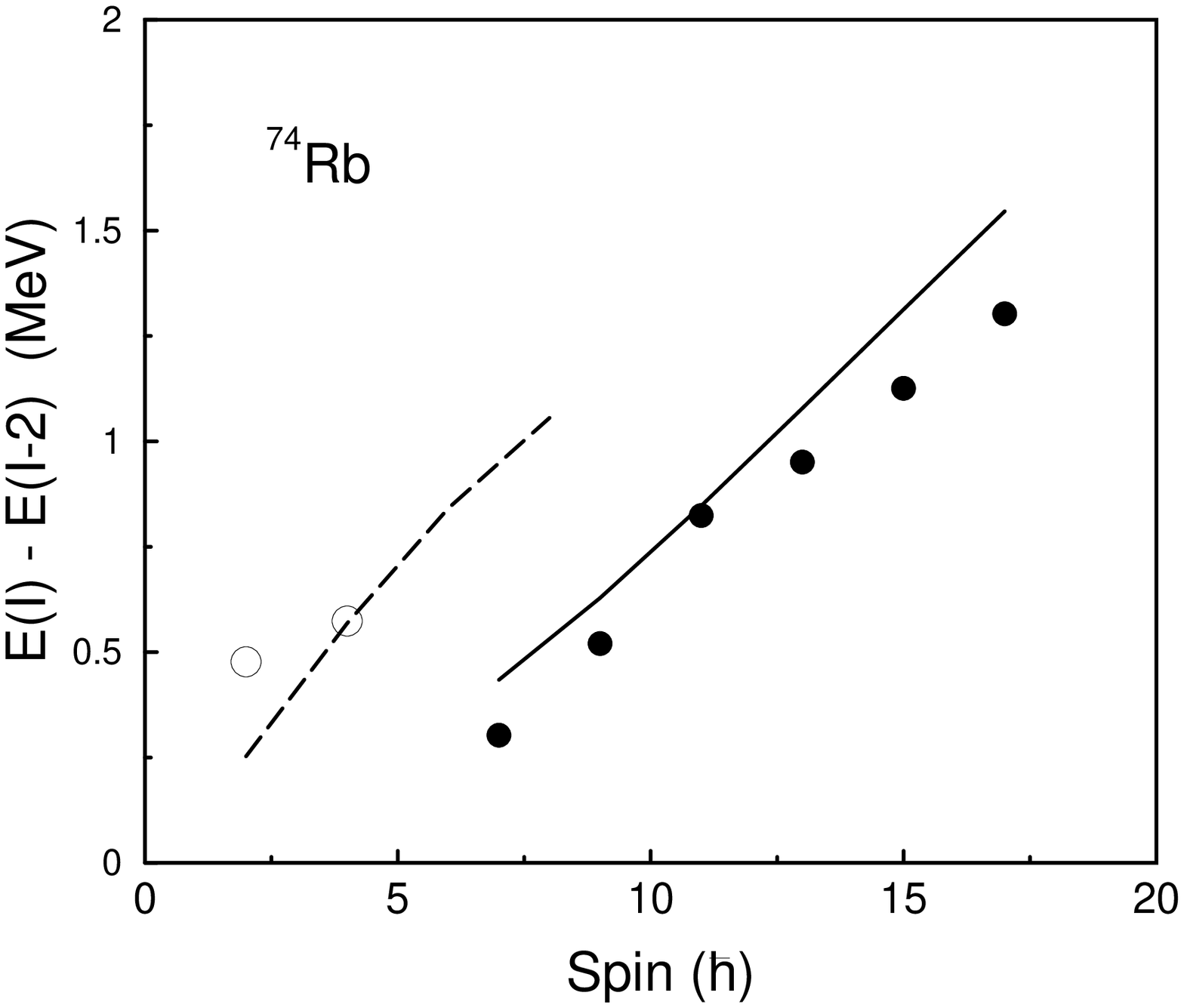}
\caption{Comparison of calculated transition energies (dashed line for 0-qp configuration and 
solid line for 2-qp configuration) with 
experimental values (open circle for the ground state band and filled circle for the excited band) 
for $^{74}$Rb 
(data from \protect\cite{74rb1}).}
\end{center}
\end{figure}


\begin{thebibliography}{999}

\bibitem{70br1} G. de Angelis, T. Martinez, A. Gadea, N. Marginean, E. Farnea, E. Maglione, S. Lenzi, W. Gelletly, C.A. Ur, D.R. Napoli, Th. Kroell, S. Lunardi, B. Rubio, M. Axiotis, D. Bazzacco, A.M. Bizzeti Sona, P.G. Bizzeti, P. Bednarczyk, A. Bracco, F. Brandolini, F. Camera, D. Curien, M. De Poli, O. Dorvaux, J. Eberth, H. Grawe, R. Menegazzo, G. Nardelli, J. Nyberg, P. Pavan, B. Quintana, C. Rossi Alvarez, P. Spolaore, T. Steinhart, I. Stefanescu, O. Thelen, and R. Venturelli, Eur. Phys. J. A12, 51 (2001).

\bibitem{70br2} D.G. Jenkins, N.S. Kelsall, C.J. Lister, D.P. Balamuth, 
M.P. Carpenter, T.A. Sienko, S.M. Fischer, R.M. Clark, P. Fallon, 
A. G\"orgen, A.O. Macchiavelli, C.E. Svensson, R. Wadsworth, W. Reviol, 
D.G. Sarantites, G.C. Ball, J. Rikovska Stone, O. Juillet, P. Van Isacker, 
A.V. Afanasjev, and S. Frauendorf, Phys. Rev. C65, 064307 (2002).

\bibitem{72br1} N. Fotiades, J.A. Cizewski, C.J. Lister, C.N. Davids, 
R.V.F. Janssens, D. Seweryniak, M.P. Carpenter, T.L. Khoo, T. Lauritsen, D. Nisius, 
P. Reiter, J. Uusitalo, I. Wiedenhover, A.O. Macchiavelli, and R.W. MacLeod,
Phys. Rev. C60, 057302 (1999).
 
\bibitem{74br1} G. Garcia-Bermudez, M.A. Cardona, A. Filevich, R.V. Ribas, 
H. Somacal, and L. Szybisz, Phys. Rev. C59, 1999 (1999).

\bibitem{74br2} J.W. Holcomb, T.D. Johnson, P.C. Womble, P.D. Cottle, 
S.L. Tabor, F.E. Durham, and S.G. Buccino, Phys. Rev. C43, 470 (1991). 

\bibitem{74rb1} D. Rudolph, C.J. Gross, J.A. Sheikh, D.D. Warner, I.G. Bearden, R.A. Cunningham, D. Foltescu, W. Gelletly, F. Hannachi, A. Harder, T.D. Johnson, A. Jungclaus, M.K. Kabadiyski, D. Kast, K.P. Lieb, H.A. Roth, T. Shizuma, J. Simpson, \"O. Skeppstedt, B.J. Varley, and M. Weiszflog, 
Phys. Rev. Lett. 76, 376 (1996).

\bibitem{76rb1} A. Harder, M.K. Kabadiyski, K.P. Lieb, D. Rudolph, C.J. Gross, 
R.A. Cunningham, F. Hannachi, J. Simpson, D.D. Warner, H.A. Roth, \"O. Skeppstedt, 
W. Gelletly, and B.J. Varley, Phys. Rev. C51, 2932 (1995). 

\bibitem{78rb1} R.A. Kaye, J. D\"oring, J.W. Holcomb, G.D. Johns, T.D. Johnson, 
M.A. Riley, G.N. Sylvan, P.C. Womble, V.A. Wood, S.L. Tabor, and J.X. Saladin,
Phys. Rev. C54, 1038 (1996). 

\bibitem{78rb2} R.A. Kaye, L.A. Riley, G.Z. Solomon, S.L. Tabor, and P. Semmes,
Phys. Rev. C58, 3228 (1998).

\bibitem{78y1} J. Uusitalo, D. Seweryniak, P.F. Mantica, J. Rikovska,
D.S. Brenner, M. Huhta, J. Greene, J.J. Ressler, B. Tomlin, C.N. Davids,
C.J. Lister, and W.B. Walters, Phys. Rev. C57, 2259 (1998).

\bibitem{80y1} D. Bucurescu, C.A. Ur, D. Bazzacco, C. Rossi-Alvarez, P. Spolaore, 
C.M. Petrache, M. Ionescu-Bujor, S. Lunardi, N.H. Medina, D.R. Napoli, M. De Poli, 
G. de Angelis, F. Brandolini, A. Gadea, P. Pavan, and G.F. Segato, Z. Phys. A352, 361 
(1995).

\bibitem{82y1} S.D. Paul, H.C. Jain, S. Chattopadhyay, M.L. Jhingan, and
J.A. Sheikh, Phys. Rev. C51, 2959 (1995).

\bibitem{rp} H. Schatz, A. Aprahamian, J. G\"orres, M. Wiescher, T. Rauscher,
J.F. Rembeges, F.-K. Thielemann, B. Pfeiffer, P. M\"oller, K.-L. Kratz, H. Herndl,
B.A. Brown and H. Rebel, Phys. Rep. 294, 167 (1998).

\bibitem{rp3} J. Garces Narro, C. Longour, P.H. Regan, B. Blank, C.J. Pearson, M. Lewitowicz, C. Miehe, W. Gelletly, D. Appelbe, L. Axelsson, A.M. Bruce, W.N. Catford, C. Chandler, R.M. Clark, D.M. Cullen, S. Czajkowski, J.M. Daugas, P. Dessagne, A. Fleury, L. Frankland, J. Giovinazzo, B. Greenhalgh, R. Grzywacz, M. Harder, K.L. Jones, N. Kelsall, T. Kszczot, R.D. Page, A.T. Reed, O. Sorlin, and R. Wadsworth, Phys. Rev. C63, 044307 (2001).

\bibitem{kr79} A.J. Kreiner and M.A.J. Mariscotti, Phys. Rev. Lett. 43, 1150 
(1979).

\bibitem{re01} R. Zheng, S. Zhu, N. Cheng, J. Wen, Phys. Rev. C64, 014313 
(2001).

\bibitem{pa1} J. D\"oring, D. Pantelica, A. Petrovici, B.R.S. Babu, J.H. Hamilton, 
J. Kormicki, Q.H. Lu, A.V. Ramayya, O.J. Tekyi-Mensah, and S.L. Tabor, Phys. Rev. C57, 
97 (1998). 

\bibitem{pa2} A. Petrovici, K.W. Schmid, and A. Faessler, Nucl. Phys. A647, 197 (1999). 

\bibitem{cau94} E. Caurier, A.P. Zuker, A. Poves, and  G. Mart\'\i nez-Pinedo,
Phys. Rev. C50 (1994) 225. 

\bibitem{ha95} K. Hara and Y. Sun, Int. J. Mod. Phys. E4, 637 (1995).

\bibitem{ha91} K. Hara and Y. Sun, Nucl. Phys. A531, 221 (1991).

\bibitem{pa01} R. Palit, J. A. Sheikh, Y. Sun, and H. C. Jain, Nucl. Phys.
A686 141 (2001).

\bibitem{SS01} Y. Sun and J.A. Sheikh, Phys. Rev. C64, 031302 (2001).

\bibitem{su94} Y. Sun and J.L. Egido, Nucl. Phys. A580, 1 (1994).  

\bibitem{su00} Y. Sun, J.-y. Zhang, M. Guidry, J. Meng, and S. Im, Phys. Rev. C62, 
021601 (2000).  

\bibitem{FS99} S. Frauendorf and J.A. Sheikh, Nucl. Phys. A645, 509 (1999).


\end{thebibliography}
\end{document}